\PassOptionsToPackage{scheme=plain, fontset=none}{ctex}

\documentclass[referee]{raa}

\usepackage{enumitem}
\usepackage{makecell}
\usepackage{graphicx,times}
\usepackage{natbib}
\usepackage[UTF8]{ctex}
\usepackage{amssymb,amsmath}
\bibpunct{(}{)}{;}{a}{}{,}
\usepackage{booktabs}
\usepackage{threeparttable}

\usepackage{url}
\usepackage[pagebackref=true]{hyperref}

\begin{document}

   \title{An Attempt to Search for Unintended Electromagnetic Radiation from Starlink Satellites with the 21 Centimeter Array: Methodology and RFI Characterization}

 \volnopage{ {\bf 20XX} Vol.\ {\bf X} No. {\bf XX}, 000--000}
   \setcounter{page}{1}

   \author{
    Xupiao Yang
    \inst{1}  
   \and 
    Qijun Zhi
   \inst{1,3}
   \and Yanbin Yang \thanks{E-mail: yangyanbin@shao.ac.cn}
   \inst{2,5,6} 
   \and Quan Guo  \thanks{E-mail: guoquan@shao.ac.cn}
   \inst{4}
   \and Juhua Gu  \thanks{E-mail: jhgu@nao.cas.cn}
   \inst{7}   
   \and Jianfeng Wang 
   \inst{8}
   \and Yan Huang
   \inst{7}
   \and Yun Yu
   \inst{2, 4}
   \and Feiyu Zhao
   \inst{2, 5}
   }

\institute{
    School of Physics and Electronic Science, Guizhou Normal University, Guiyang 550001, China \\
    \and 
   Shanghai Astronomical Observatory, Chinese Academy of Sciences, 80 Nandan Road, Shanghai 200030, China\\
   \and
   School of Physics Science, Guizhou University, Guiyang 550001, China \\
   \and 
   State Key Laboratory of Radio Astronomy and Technology,  Shanghai Astronomical Observatory, Chinese Academy of Sciences, 80 Nandan Road, Shanghai 200030, China\\
   \and
   University of Chinese Academy of Sciences, No.1 Yanqihu East Road, Beijing 101408, China\\
   \and
   School of Physical Science and Technology, ShanghaiTech University, Shanghai 201210, China\\
   \and 
    State Key Laboratory of Radio Astronomy and Technology, National Astronomical Observatories, Chinese Academy of Sciences, 20A Datun Road, Beijing 100101, China 
   \and 
   National Astronomical Observatories, Chinese Academy of Sciences, 20A Datun Road, Beijing 100101, China
   }

\vs \no
   {\small Received 20XX Month Day; accepted 20XX Month Day}

\abstract{The rapid expansion of low-Earth-orbit (LEO) megaconstellations introduces new risks to radio astronomy from unintended electromagnetic radiation (UEMR). In this work, we present an attempt to search for UEMR from Starlink
satellites using the 21 Centimeter Array (21CMA). Because the sensitivity of a single pod observation is limited, we focus on developing a robust observing and detection
pipeline. Using Two-Line Element (TLE) data, we predict satellite transit times to
guide the observations, and we define entry into the field of view (FoV) as an apparent declination greater than $85^{\circ}$ with respect to the 21CMA. We analyze the system equivalent flux density (SEFD) and the resulting single-pod sensitivity limits,
which explain the detection of emission originating from the ORBCOMM satellites, rather than any detectable broadband UEMR in our dynamic spectra. To validate the methodology, we developed a Python package, orbdemod, to demodulate ORBCOMM downlink signals in our data. The recovered satellite ID agrees with the satellite predicted by our maximum-declination analysis, thereby validating
the accuracy of our transit prediction and identification framework. Furthermore, via
modulation power spectrum analysis, we show that the impulsive broadband bursts are produced by power line arcing near the array rather than by satellite UEMR.
}

   \authorrunning{Xupiao Yang et al. }            
   \titlerunning{An Attempt to Search for UEMR with the 21CMA}  
   \maketitle

%
\section{Introduction}           
\label{sect:intro}

In recent years, the number of LEO satellites has grown at an unprecedented pace, markedly intensifying global occupation of the radio frequency spectrum. Major constellation operators include SpaceX (Starlink, USA), Amazon (Kuiper, USA), Eutelsat (OneWeb, UK), Shanghai Spacecom Satellite Technology (Qianfan, China), China Satellite Network Group (Guowang, China), and Russia's Sfera constellation (\citealt{Grigg2025AAgrowing}). At present, Starlink alone has launched more than 10,000 spacecraft (over 8,000 on orbit) and plans to expand to about 42,000 within the next decade. Such a massive satellite population will have far-reaching consequences for the radio frequency environment of ground-based astronomy.

Starlink satellites have been observed to exhibit unintended electromagnetic radiation (UEMR) at low frequencies, well outside their nominal communication bands above 10 GHz. LOFAR has reported the detection of both broadband and narrowband UEMR events in the frequency ranges of 40--70～MHz and 110--188～MHz, with maximum spectral power flux densities on the order of $10^3$ Jy (\citealt{Vruno2023AA, bassa2024A&A}). In addition, the SKA-Low prototype station EDA2 independently identified intended electromagnetic radiation from Starlink satellites at 137.5 MHz, attributed to transmissions from the ‘Swarm’ transmitters, as well as UEMR at 159.4 MHz (\citealt{grigg2023A&A}). \citet{zhangx2025AA} used NenuFAR to investigate the impact of Starlink satellites at frequencies below 100～MHz, finding broadband UEMR predominantly concentrated in the 54--66~MHz. The emission exhibited intensities exceeding 500 Jy and was highly polarized, with significant differences in emission properties observed between satellite generations.

For experiments aiming to detect the ultra-faint redshifted 21-cm signal from neutral hydrogen during the Epoch of Reionization (EoR), unintended electromagnetic radiation (UEMR) from satellites has emerged as a significant and increasingly non-negligible challenge. First, the intensity of satellite UEMR can be orders of magnitude stronger than the target 21-cm signal, severely contaminating the data and potentially compromising detection efforts. Second, although radio telescopes are typically located in remote regions to minimize terrestrial interference, contamination from satellites overhead is unavoidable. Systematic observations and quantitative characterization of satellite UEMR are therefore essential for developing effective mitigation strategies. Such studies also provide a scientific foundation for informed dialogue and strengthened collaboration with satellite operators.

To date, studies of Starlink UEMR have been conducted primarily in the Western and Southern Hemispheres, while observational reports from East Asia remain limited. Given the rapid expansion of large satellite constellations, including those deployed by Chinese institutions, systematic investigations of UEMR using the 21CMA are both timely and necessary. Such efforts, beginning with the Starlink constellation and extending to the Qianfan (G60) and other satellite systems, will provide important regional measurements and broaden the global understanding of satellite-induced interference.

The 21CMA is currently advancing toward the implementation of multi-beam coherent digital beamforming techniques, in line with developments in modern low-frequency aperture arrays. Such capabilities are essential for enhancing sensitivity, improving interference discrimination, and enabling flexible beam steering for next-generation experiments. In this work, however, observations are conducted using a single pod configuration. Owing to the limited sensitivity of this setup, our primary objective is not to achieve the deepest possible detection threshold, but rather to establish, implement, and validate a robust observational strategy for satellite UEMR studies.Section~\ref{sect:Obs} introduces our observational strategy. 

In Section~\ref{sensitivity} and Section~\ref{UEMR}, we explain the absence of clearly identifiable broadband UEMR through an analysis of the system equivalent flux density (SEFD) and the resulting sensitivity limits. In Section~\ref{RFI}, power line arcing RFI is investigated and consequently excluded as a UEMR candidate, and In Section~\ref{ORBCOMM}, we verify the validity of the observational strategy by demodulating the downlink transmissions of ORBCOMM satellites. We provide our discussion and conclusions in Section~\ref{conclusion}.

\section{Observation}

\label{sect:Obs}
The 21 CentiMeter Array (21CMA), located in Ulastai, Xinjiang, China, is a ground-based radio interferometer that serves as one of the Square Kilometre Array (SKA) pathfinders. The array is arranged in two perpendicular arms—east-west (6.1 km) and north-south (4 km)—comprising a total of 81 pods \cite{zheng2016ApJ}. Each pod contains 127 log-periodic antennas, bringing the total antenna count to 10,287. The antennas are fixed and pointed towards the North Celestial Pole (NCP), with a field of view (FoV) covering a circular region of approximately $5^{\circ}$ radius centered on the NCP. The 21CMA operates in the frequency range of 50–200 MHz, and the effective area for a single pod is about $218\text{ m}^2$ in the frequency range of 75–200 MHz.

This study utilizes raw voltage data from E8 (the 8th pod in the east arm of the array). The data is sampled at a rate of 480 Msps in int16 format, corresponding to a Nyquist bandwidth of 240 MHz. Owing to the bandpass characteristics of the 21CMA system’s filters, the effective observational frequency range is constrained to 50–200 MHz. 

To optimize data storage efficiency and maximize the acquisition of valid satellite signal samples, we adopt a targeted observation strategy based on satellite ephemerides. Observation windows were pre-calculated according to predicted satellite passes, with data recording triggered only during these transits. Compared to continuous 24-hour monitoring, this approach significantly increases the fraction of Starlink satellite transit events captured within a limited data volume.

We retrieved the Two Line Elements (TLEs) data for all active satellites on the day of observation from the CelesTrak \footnote{https://celestrak.org/} website to calculate satellite trajectories. Using the open-source \texttt{Skyfield} \footnote{https://rhodesmill.org/skyfield/} Python library (\citealt{Rhodes}), we integrated the TLEs with the geographic coordinates of 21CMA (42.9333°N, 86.6833°E, elevation 2650 m) to determine satellite positions within the specified observation time. Given that the 21CMA is fixed toward the NCP with a $5^{\circ}$ radius field of view, an apparent declination greater than $85^{\circ}$ was adopted as the criterion for a satellite transit event. This approach allowed us to accurately predict the ingress and egress times of various satellites and to select periods with high-density Starlink passes as our primary observation windows. Furthermore, since the orbital altitudes of LEO satellites significantly exceed the physical spacing between the 21CMA pods, and considering that a typical Starlink transit lasts approximately 20 seconds, the timing delays caused by the slight spatial offset of an individual pod relative to the array center are negligible at the precision required for our analysis.

Ultimately, following the aforementioned methodology, we prioritized observation windows that included satellites studied in prior work (e.g., Starlink-3618 and Starlink-3645 \citep{Vruno2023AA}), as well as the newly launched Chinese Qianfan (G60) constellation. In total, five observation periods were selected on April 15, 2025 (UTC), accumulating approximately 2.2 TB of raw data. This dataset encompasses Starlink satellites ranging from version 1.0 to the v2-mini Direct-to-Cell (D2C). Detailed information regarding the observation windows and representative satellites is summarized in Table~\ref{tab:observation-time}. Representative data samples from these observations are publicly available via the Zenodo repository\footnote{10.5281/zenodo.18213739}.

\begin{table}[h]
\centering
\begin{threeparttable}
\caption{Summary of the five observation sessions on April 15, 2025 (UTC)}
\label{tab:observation-time}
\begin{tabular}{lcc}
\hline
Obs. ID & Time Windows (UTC) & Representative Satellites\\

\hline
1 & 11:39:52 -- 11:51:06 &\makecell{STARLINK-1966 \\ STARLINK-3618 \\ QIANFAN-26} \\ \midrule
2 & 12:59:53 -- 13:09:41 &\makecell{STARLINK-1433 \\ STARLINK-3645 \\ STARLINK-5499} \\ \midrule
3 & 13:59:56 -- 14:14:34 &\makecell{STARLINK-11072 [DTC]\tnote{*} \\ STARLINK-31333 \\ QIANFAN-13} \\ \midrule
4 & 16:35:53 -- 16:38:44 &\makecell{STARLINK-1966 \\ STARLINK-4189 \\ STARLINK-4198} \\ \midrule
5 & 23:34:52 -- 23:38:00 &\makecell{STARLINK-32153 \\ STARLINK-11237 [DTC]\tnote{*} \\ QIANFAN-36} \\ 
\hline
\textbf{Total Satellite Passes} &  --- & \textbf{58} \\
\hline\noalign{\smallskip}
\end{tabular}
\begin{tablenotes}
    \footnotesize
    \item[*] DTC: Direct-to-Cell capability.
\end{tablenotes}
\end{threeparttable}
\end{table}

\section{Analysis and Results}
\subsection{Sensitivity Analysis}
\label{sensitivity}
We performed a theoretical assessment of the observational sensitivity based on the system parameters of a single 21CMA pod. The system equivalent flux density (SEFD) is defined as:

\begin{equation}
\rm{SEFD} = \frac{2k_BT_{sys}}{A_{eff}}
\end{equation}

where $k_B$ is the Boltzmann constant, $A_{\rm eff}$ is the effective collecting area of a single pod, and $T_{\rm sys}$ is the system temperature. $T_{\rm sys}$ incorporates contributions from the sky background noise, receiver noise, and system losses. At 150 MHz, the effective collecting area of a 21CMA pod is approximately $218\ \rm {m^2}$. In this frequency range, the system temperature is dominated by the sky background noise, primarily originating from diffuse galactic synchrotron emission. 

Using the Global Sky Model (GSM2016) \citep{zheng2016GSM} via the \texttt{PyGDSM}  \footnote{https://github.com/telegraphic/pygdsm} Python library, we estimated the sky brightness temperature within a circular region of $5^{\circ}$ radius centered on the NCP. The average brightness temperature and its standard deviation were found to be $253.2 \pm 14.9\, \text{K}$ .

Given that the receiver noise temperature for the 21CMA system is approximately $50 \, \text{K}$ \citep{huang2016RAA}, and accounting for additional contributions from ground radiation picked up by sidelobes and losses within the cables and electronics, we conservatively adopted $T_{\rm sys}  = 350\, \text{K}$ as the representative system temperature at $150 \,  \text{MHz}$. This yields an estimated SEFD for a single 21CMA pod of approximately $4.43 \, \text{kJy}$.

Based on the calculated SEFD, the theoretical sensitivity of the single pod system can be derived using the radiometer equation. Since the 21CMA employs single polarization antennas, the sensitivity $\Delta S$ is expressed as:
\begin{equation}
    \Delta S = \frac{\rm SEFD}{\sqrt{\Delta \nu \tau}}
\end{equation}

where $\Delta \nu$ denotes the frequency resolution and $\tau$ represents the integration time. In the initial data processing, we performed autocorrelation on the raw voltage data with an FFT size of 16,384 points, yielding a frequency resolution of $29.3 \,\text{kHz}$ and a temporal resolution of $34.13\,\mu\text{s}$. At this raw resolution, the theoretical sensitivity $\Delta S$ is approximately $4.43~\text{kJy}$.

In Figure 6 of \citet{Vruno2023AA}, a clear dynamic spectrum of Starlink's broadband UEMR is presented with a temporal resolution of $41\,\text{ms}$ and a frequency resolution of $0.195\,\text{MHz}$. Given the characteristics of the 21CMA data, we adopted a comparable resolution strategy: a factor of 1,024 average was applied in the time domain to reach a $35\,\text{ms}$ resolution, and an 8-channel average was applied in the frequency domain to reach a $0.234\,\text{MHz}$ resolution. Under these specific parameters, we achieved a theoretical sensitivity of $48.95\,\text{Jy}$. Assuming a $3\sigma$ confidence level, the minimum detectable flux density for satellite UEMR in this study is established at $146.85\,\text{Jy}$. 

\subsection{UEMR Search}
\label{UEMR}
\begin{figure}[hbt!]
   \centering
   \includegraphics[width=0.8\columnwidth, angle=0]{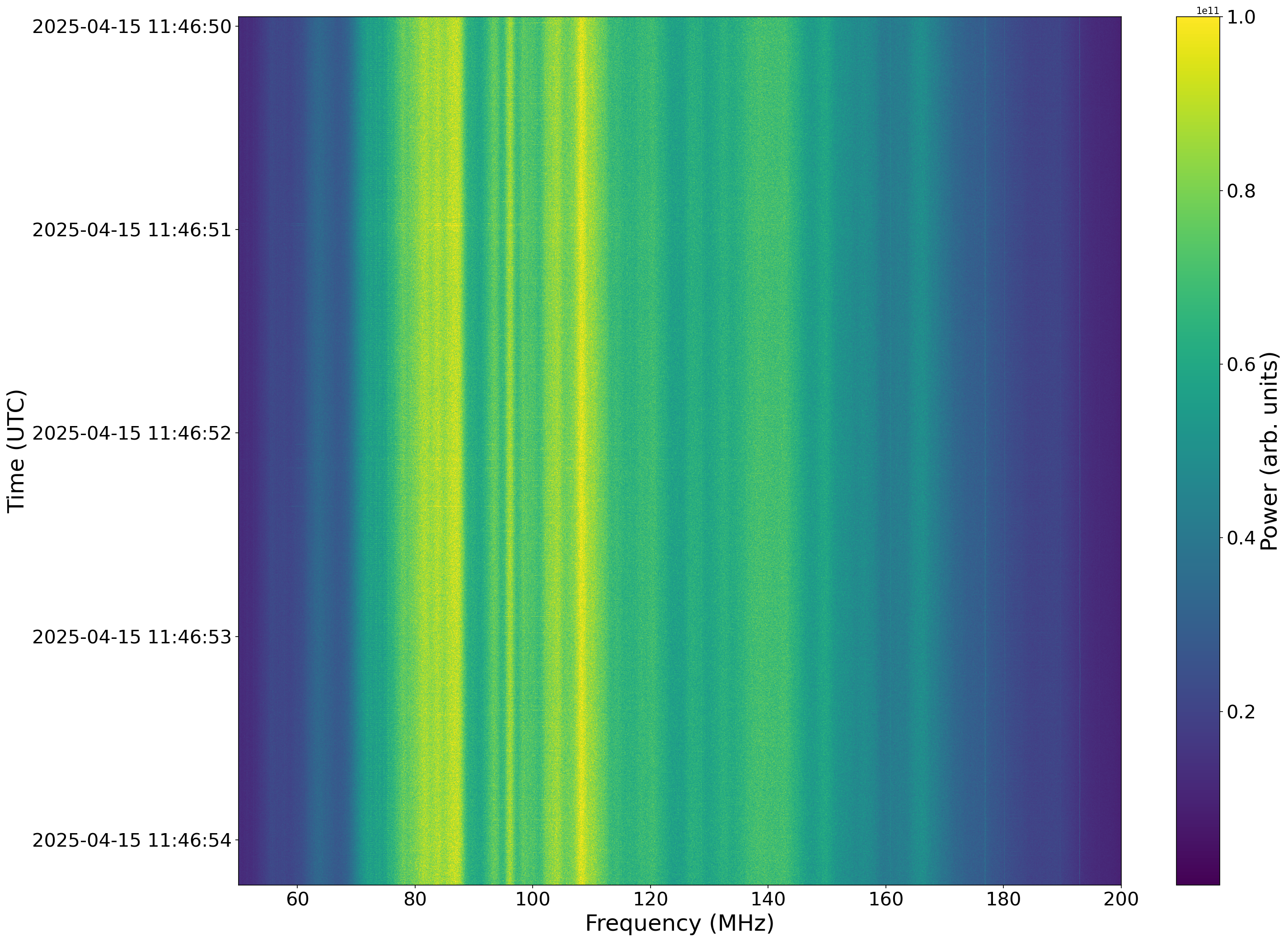}
   \caption{An example of the dynamic spectra of 21CMA single pod E8. The dynamic spectra is averaged by a factor of 1024 in the time domain, yielding a time resolution of $35~\mathrm{ms}$, and by 8 channels in the fr equency domain, resulting in a frequency resolution of $0.234~\mathrm{MHz}$. Within the frequency range of 110–188 MHz, there is no obvious broadband UEMR.}
   \label{21CMAwaterfall}
   \end{figure}

Following the averaging parameters described in Section~\ref{sensitivity}, we generated a sequence of dynamic spectra, each with a duration of approximately $4.27\,\text{s}$. Figure~\ref{21CMAwaterfall} presents a representative example. The dynamic spectrum corresponds to a selected time interval within the period during which Starlink-3618 (NORAD ID: 51998) transited the 21CMA’s field of view, from 2025-04-15 11:46:44 to 11:47:03 (UTC). The UEMR of this satellite has been illustrated in Figure 7 of \citet{Vruno2023AA}. However, in our Figure~\ref{21CMAwaterfall}, no broadband UEMR is found within the frequency range of 110–188 MHz.

Actually, manual inspection revealed no clearly identifiable or unambiguous broadband UEMR features in our dynamic spectra during any of the predicted satellite transit time windows. Based on the single pod sensitivity analysis, we attribute this null result to the intrinsic sensitivity limitations of the current observation setup.

In the 100-200 MHz range, previous studies by \citet{Vruno2023AA} and \citet{bassa2024A&A} reported that the spectral power flux density of Starlink's broadband UEMR typically ranges from a few to 100 Jy. However, our calculated theoretical sensitivity at 150 MHz is $\Delta S = 48.95$ Jy, corresponding to a $3\sigma$ detection threshold of 146.85 Jy. This threshold significantly exceeds the reported signal intensities, even without considering additional base-level fluctuations caused by RFI, thereby explaining the invisibility of these faint features.

For frequencies below 100 MHz, although radiation intensity increases (with UEMR exceeding 500 Jy in the 54–66 MHz band as noted by \citet{bassa2024A&A} and \citet{zhangx2025AA}), the detection environment becomes increasingly hostile. As frequency $\nu$ decreases, the diffuse galactic synchrotron emission intensifies, following a $\nu^{-2.55}$ power law. Meanwhile, due to aperture saturation effects, the effective area $A_{\rm eff}$ of a single pod increases less steeply than the theoretical $\nu^{-2}$ scaling, following a more conservative $\nu^{-1.5}$ dependence. Taking 60 MHz as an example, while the sky temperature increases by a factor of ~10.35 compared to 150 MHz, the effective area only expands by a factor of ~3.95. This results in a degraded sensitivity of approximately 128.26 Jy ($\Delta S$), or 384.78 Jy at a $3\sigma$ level. Furthermore, since the 54–66 MHz range is adjacent to the 50 MHz high-pass limit of the 21CMA filters, the filter roll-off effect and subsequent gain attenuation pose additional challenges for identifying signals even at the 500 Jy level.

Once tied-array beamforming (TAB) is implemented, it will drastically enhance our detection capabilities. Following \citet{kondratiev2016}, the effective area of a tied-array beam can be approximated as:
\begin{equation}
    A_{\rm eff}^{\rm tab} = \eta_{\rm active}N^{0.85}A_{\rm eff}^{\rm station}
\end{equation}

Adopting the same configuration as \citet{Vruno2023AA}, we set the active dipole fraction to $\eta_{\rm active} = 0.95$ and the number of pods whose signals are coherently combined (forming the tied-array beam) to $N = 12$. The effective aperture area of a single pod (station) is denoted by $A_{\rm eff}^{\rm station}$. The use of 12 coherent pods will improve the system sensitivity to approximately 6.23 Jy. Such an upgrade will provide the 21CMA with the necessary sensitivity to detect broadband Starlink UEMR across the 100–200 MHz range.

\subsection{RFI Environment}
\label{RFI}
\begin{figure}[htbp]
    \centering
    \includegraphics[width=0.85\linewidth]{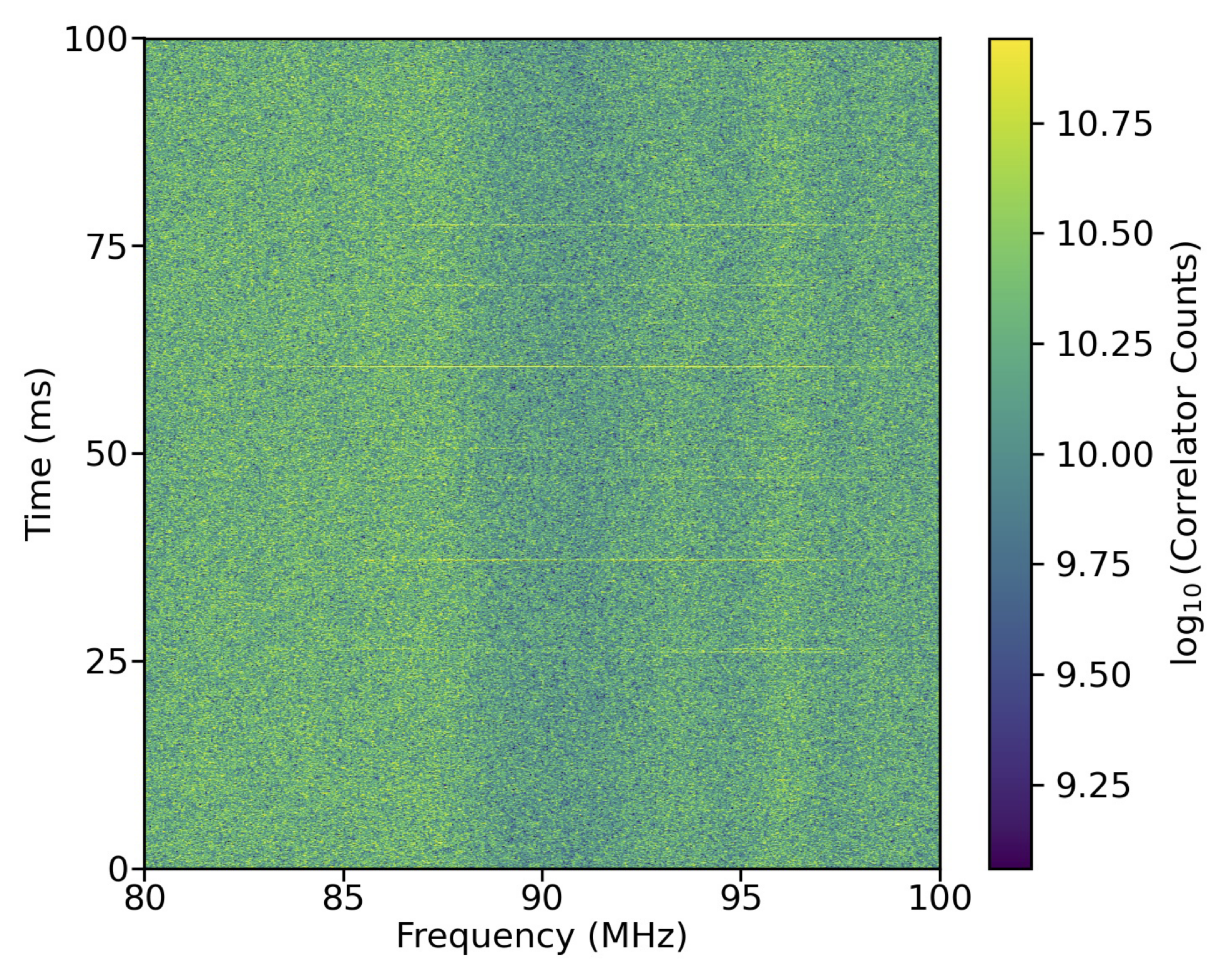}
    \caption{
    Several examples of power line arcing RFI. The horizontal axis spans 80–100~MHz with a frequency resolution of 29.3~kHz, while the vertical axis covers a duration of 100~ms with a time resolution of $34~\mu\text{s}$.
    }
    \label{broadbandRFI}
\end{figure}

Although no definitive satellite UEMR is identified in the averaged dynamic spectra,
several noteworthy RFI is observed. Both the averaged and raw dynamic spectra reveal recurrent transient broadband signals with instantaneous bandwidths spanning from tens to over 100 MHz, particularly prominent within the 50–100 MHz range in raw dynamic spectra. As illustrated in Figure~\ref{broadbandRFI}, these events are highly impulsive, typically lasting only a single temporal resolution element ($34~\mu\text{s}$). Initially, these were considered potential candidates for satellite UEMR due to their broadband nature. However, subsequent temporal correlation analyses indicated that their occurrence patterns do not align with predicted satellite transits. Further investigation leads us to conclude that these signals originate from power line arcing RFI generated by nearby high-voltage transmission infrastructure.

To balance statistical robustness with computational efficiency, we performed modulation spectrum analysis on the raw dynamic spectra. The dataset was divided into intervals of approximately $128~\text{s}$ each (corresponding to 15 consecutive data files). For each part, the power within the $60\text{-}120~\text{MHz}$ range was averaged to generate a time series, which was then subjected to a secondary FFT to extract periodic modulation features of the signal intensity.

\begin{figure}[htbp]
    \centering
    \includegraphics[width=0.65\linewidth]{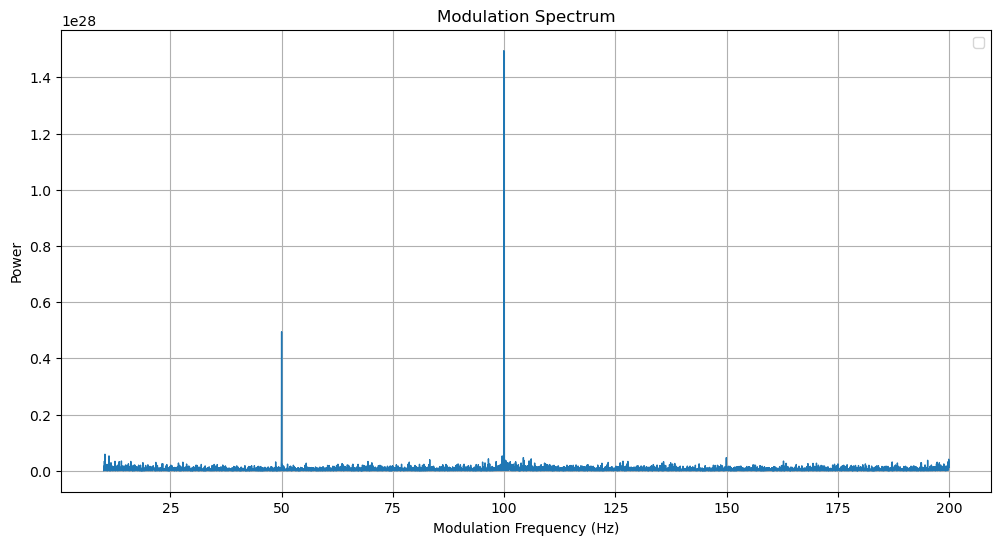}
    \caption{ An example of modulation power spectrum of the integrated 60–120 MHz band. A dominant narrowband peak at $100~\text{Hz}$ and a secondary harmonic at $50~\text{Hz}$ are clearly identified. These features are characteristic fingerprints of power line arcing.
.
    }
    \label{mod_spec}
\end{figure}

Figure～\ref{mod_spec} illustrates a representative modulation power spectrum from one such interval. A dominant, narrowband peak is clearly visible at $100~\text{Hz}$, accompanied by a secondary harmonic at $50~\text{Hz}$. This $100~\text{Hz}$ modulation signature is in excellent agreement with the physical mechanism of power line arcing, which typically occurs during both the positive and negative phases of the $50~\text{Hz}$ AC cycle. The consistent detection of this feature across multiple processed segments effectively excludes these transient broadband signals as potential satellite UEMR candidates, confirming their origin as environmental RFI from nearby high-voltage power infrastructure.

RFI remains a persistent challenge for low frequency radio astronomy, exhibiting diverse morphologies and originating predominantly from anthropogenic activities. Previous characterization of the RFI environment surrounding the 21CMA site by \citet{huang2016RAA} identified five major sources: civil aviation communications, ORBCOMM constellation broadcasts, walkies-talkies from local train communications, FM/AM broadcasts, and computer noise. In recent years, due to infrastructure development near the 21CMA site, the telescope has inevitably encountered a more complex electromagnetic environment. Our identification of power line interference signatures via modulation spectrum analysis aligns perfectly with the distribution of high-voltage transmission lines currently present around the array.

In terms of RFI detection algorithms, \citet{yang2025PASA} applied HQ-SAM, an optimized version of the Segment Anything Model (SAM), to RFI identification. By comparing it with the SumThreshold algorithm, they demonstrated the feasibility of performing automated RFI recognition without the need for task-specific fine-tuning. Building upon this, we introduce the latest SAM 2\footnote{https://ai.meta.com/research/sam2/}, which extends the Segment Anything framework from single-image segmentation to a model that maintains consistency across both images and videos. SAM 2 introduces a memory-based temporal module in its architecture: when processing a time sequence, the model maintains an explicit target state—including the mask, positional information, and prompt history—and iteratively updates this state in subsequent frames\citep{Ravi2024arXiv}. As a direct consequence, the model can reuse the context from the previous frame in the inference of the next frame, rather than treating each frame as a completely "new image" that must be segmented again from scratch. This "streaming memory" inference has been shown to improve mask consistency in video scenarios.

The capabilities of SAM 2 are highly compatible with our scientific objectives. Broadband UEMR from LEO satellites typically exhibits short durations. According to the TABs observations of \citet{Vruno2023AA}, these events often persist for only a few seconds. When treating consecutive spectra as a temporal sequence, UEMR manifests as a transient structure appearing and disappearing within the time domain. By leveraging the memory module of SAM 2, the model can perceive the temporal onset and offset of signals, enabling precise labeling and identification.

\begin{figure}[htbp]
    \centering
    \begin{subfigure}{0.49\linewidth}
        \centering
        \includegraphics[width=\linewidth]{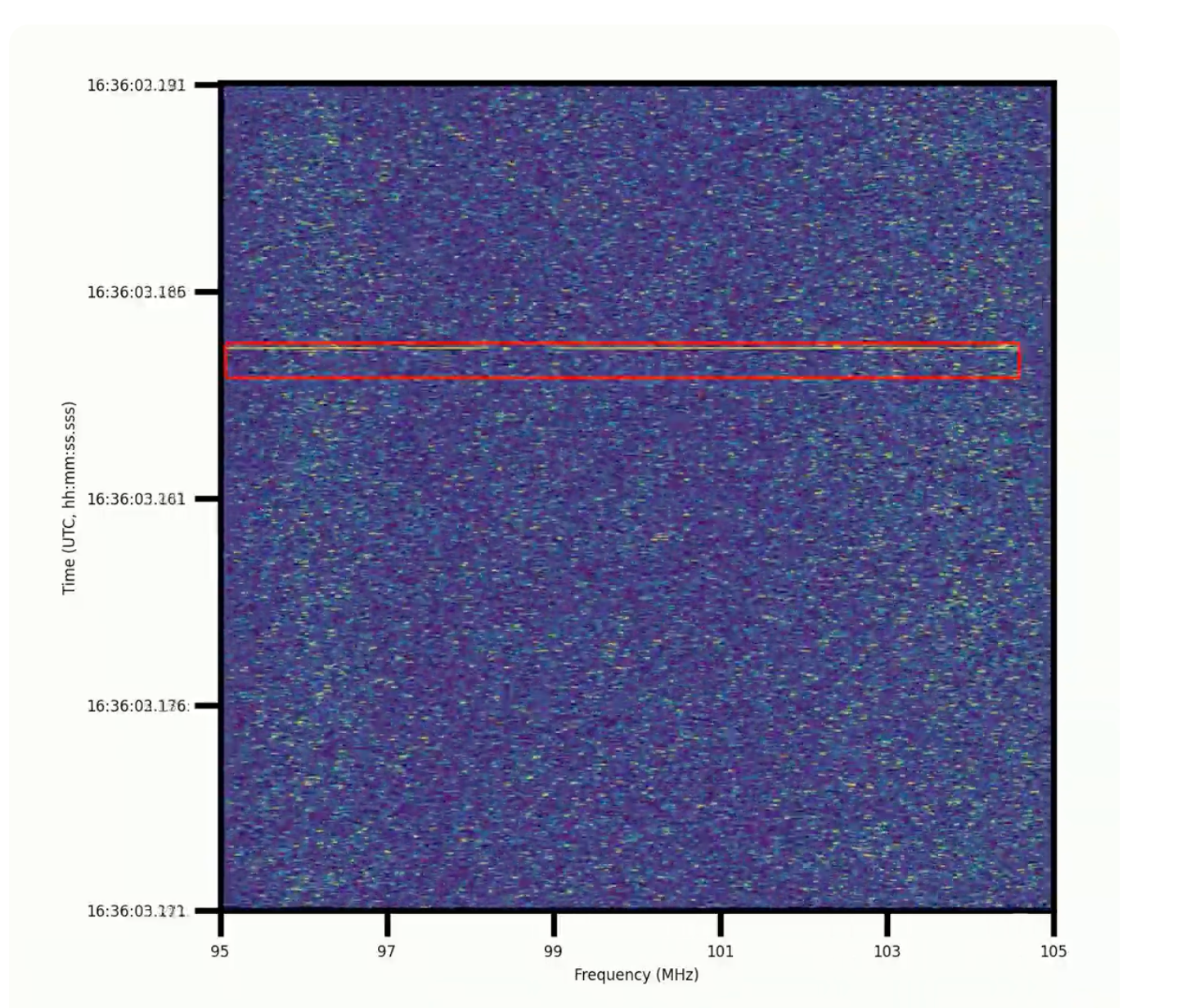}

    \end{subfigure}
    \hfill
    \begin{subfigure}{0.49\linewidth}
        \centering
        \includegraphics[width=\linewidth]{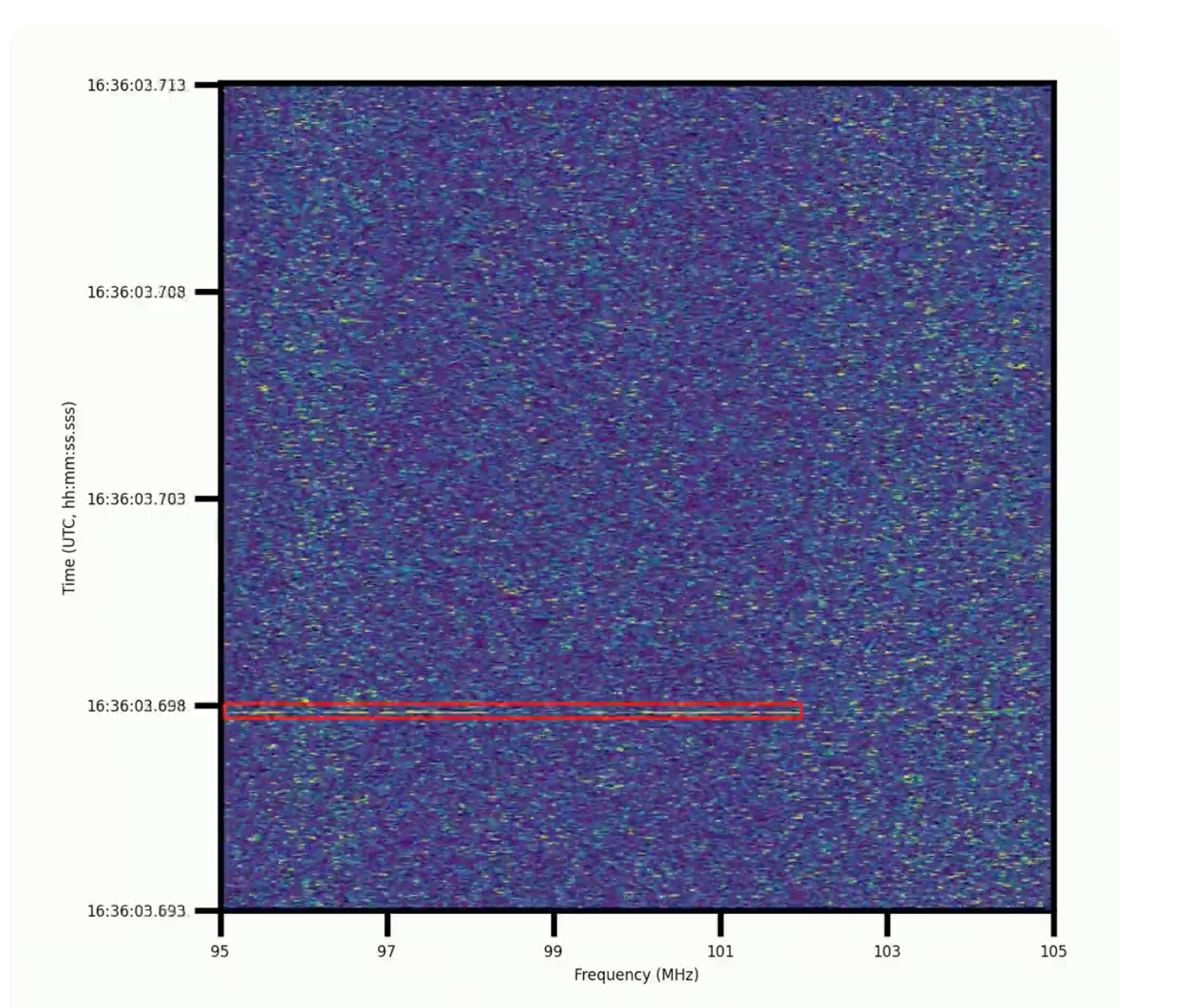}

    \end{subfigure}
    \caption{Examples of successful recognitions of power line arcing in spectra sequences using SAM 2. For a more detailed exhibition, please refer to our video.
}
    
    \label{sam}
\end{figure}

Since definitive satellite UEMR was not detected in this single pod study, we employed power line arcing RFI as a proxy to validate the algorithm's performance. We generated a series of consecutive spectra sequences with a temporal window of $0.02~\text{s}$ (resolution: $34~\mu\text{s}$) and a bandwidth of $10~\text{MHz}$ (resolution: $29.3~\text{kHz}$). The results demonstrate that SAM 2 effectively identifies the emergence of impulsive arcing  RFI across successive frames. This validation highlights the model’s capability to track non-stationary broadband features in high-cadence radio data. A demonstration video of this tracking process is available at: \url{https://youtu.be/xsp4cEdqOzw}. 
Figure～\ref{sam} presents two examples of successful detections of sudden power line arcing events in our spectra sequences, with the identified RFI marked by red boxes.

\subsection{ORBCOMM Demodulator}
\label{ORBCOMM}
\begin{figure}[hbt!]
   \centering
   \includegraphics[width=0.8\columnwidth, angle=0]{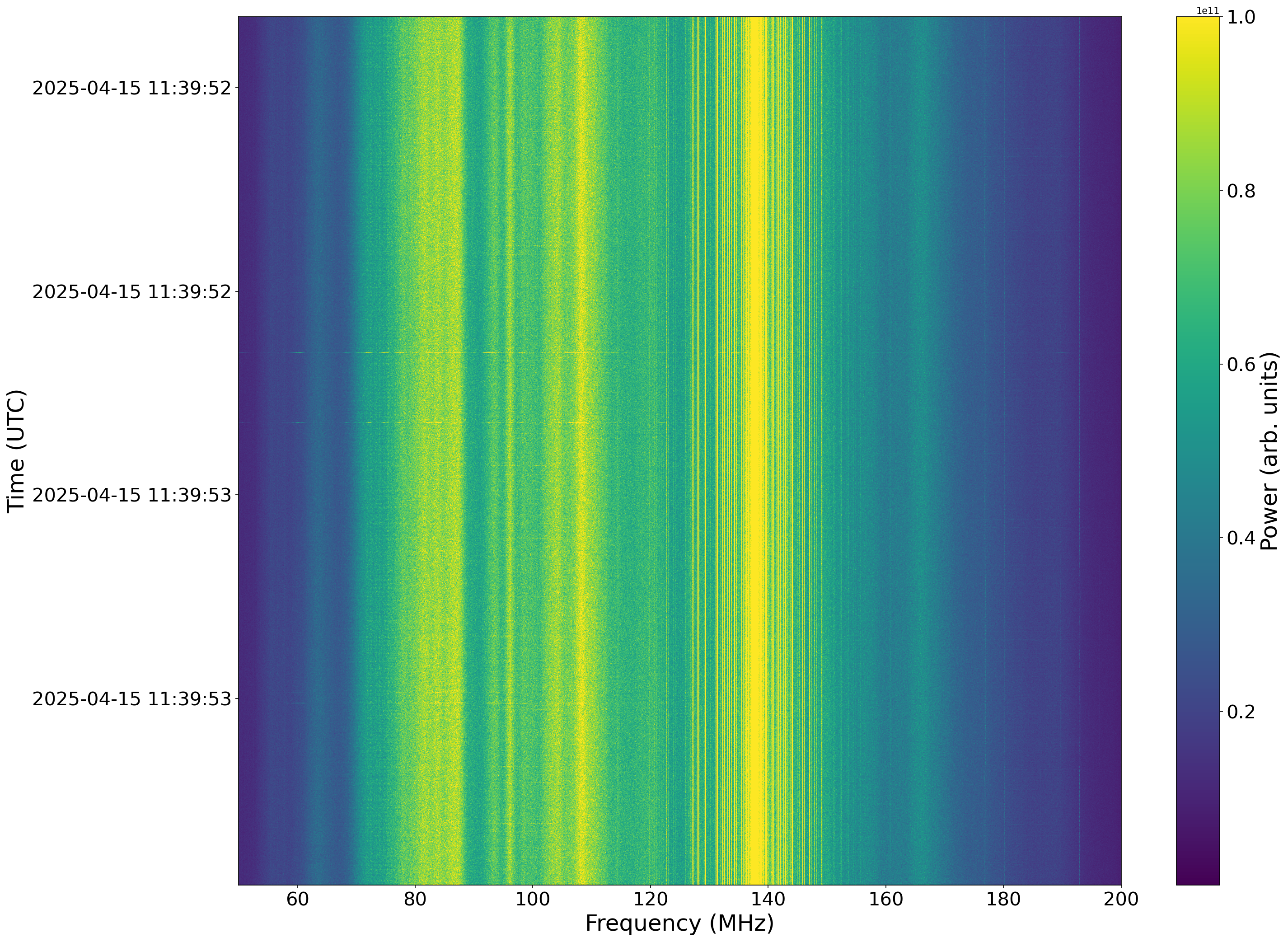}
   \caption{Dynamic spectrum recorded between 11:39:52 and 11:39:53 (UTC) on 2025-04-15. A cluster of intense narrowband RFI is clearly visible near $137~\text{MHz}$, identified as downlink transmissions from ORBCOMM satellites.}
   \label{ORBCOMM_spectrum}
   \end{figure}

In addition to the power line arcing, we identified ORBCOMM satellites downlink transmissions in the dynamic spectra, which were previously reported by \citet{huang2016RAA}. As shown in Figure~\ref{ORBCOMM_spectrum}, representing the interval around 11:39:52 to 11:39:53 (UTC) on 2025-04-15, a significant amount of narrowband RFI is prominent around $137~\text{MHz}$. Using the satellite transit prediction method described in Section~\ref{sect:Obs}, we calculated the maximum value of the apparent declination of all ORBCOMM satellites relative to 21CMA within a specific time range.

Within the time frame corresponding to Figure~\ref{ORBCOMM_spectrum}, only four ORBCOMM satellites or debris were above the horizon (the detailed information is shown in Table~\ref{tab:orbcomm}). Although none of these four objects entered the 21CMA FoV based on our criterion of declination $> 85^\circ$, the declination of ORBCOMM FM 108 was significantly higher than that of the other three. Given its relatively high declination, we have reason suspect that the $137~\text{MHz}$ signals observed in the spectrum originated primarily from this satellite, captured by the antenna through its side lobes.

\begin{table}[h]
\centering
\begin{threeparttable}
\caption{Detailed information on four ORBCOMM satellites or debris above the horizon between 11:39:52 and 11:39:53 (UTC) on 2025 April 15}
\label{tab:orbcomm}
\begin{tabular}{lcc}
\hline
SATNAME & NORAD ID & Apparent Declination\\

\hline
ORBCOMM FM 9 & 25116 &$16.5233^\circ$ \\ \midrule
ORBCOMM FM 35 & 25985 & $19.1031^\circ$\\ \midrule
ORBCOMM FM 108 & 41187 & $63.9032^\circ$ \\  \midrule
ORBCOMM FM 5 DEB & 51448 &$15.7749^\circ$ \\ \midrule

\end{tabular}

\end{threeparttable}
\end{table}

To verify our hypothesis, we demodulated the signal at $137.46~\text{MHz}$. We developed a Python package, \texttt{orbdemod} \footnote{https://github.com/JinYi1108/Orbcomm\_Demodulator} (\citealt{yang_2026_18214212}), specifically designed for the demodulation tasks. The ORBCOMM satellite downlink utilizes Symmetrical Differential Phase Shift Keying (SDPSK) modulation. At the receiver, the original digital signal is recovered by comparing the phase difference between two consecutive symbols, where phase shifts of $+90^\circ$ and $-90^\circ$ correspond to the encoded bits '1' and '0', respectively. Therefore, the absolute phase of the SDPSK signal occupies four discrete states in the IQ constellation diagram, namely $0^\circ$, $90^\circ$, $180^\circ$, and $270^\circ$.

Our software, \texttt{orbdemod}, implements the downlink demodulation through a pipeline consisting of digital down-conversion (DDC), carrier error recovery, root-raised-cosine (RRC) matched filtering, symbol timing recovery, costas loop–based phase recovery, differential decoding, minor frame synchronization, and finally, Fletcher Checksum verification and single-bit error correction.

\begin{figure}[hbt!]
   \centering
   \includegraphics[width=0.8\columnwidth, angle=0]{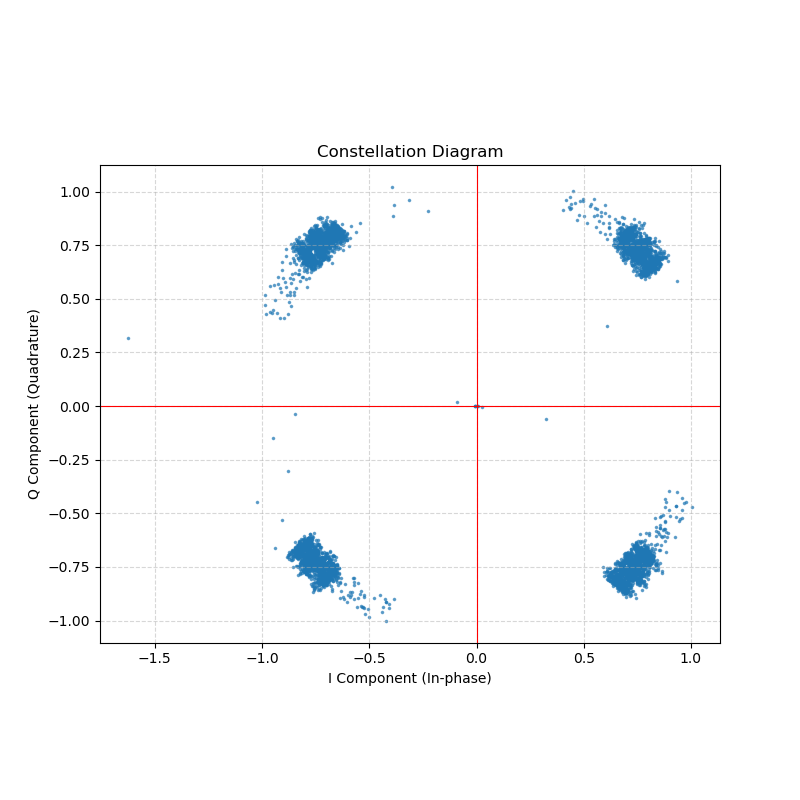}
   \caption{IQ constellation diagram of the demodulated ORBCOMM signal after Costas-loop-based carrier phase recovery, showing four tightly clustered points separated by $90^\circ$, consistent with SDPSK modulation used in the ORBCOMM downlink.}
   \label{ORBCOMM_costas}
   \end{figure}

Figure~\ref{ORBCOMM_costas} presents the IQ constellation diagram obtained from the raw voltage data (shown in Figure~\ref{ORBCOMM_spectrum}) after processing through Costas loop-based phase recovery. Four distinct clusters of sample points, separated by $90^\circ$ intervals, are clearly observable. Following this stage, the date underwent subsequent operations such as differential decoding to yield the final hexadecimal decoded output. The detailed results of the decoded packets are available at: \url{https://github.com/JinYi1108/Orbcomm_Demodulator/blob/main/examples/orbdemod_packets_iq_0.txt}.

The synchronization packet of the ORBCOMM downlink data begins with the hexadecimal sequence \texttt{65A8F9}, followed by the spacecraft identification (ID) at the fourth word. Our demodulated synchronization sequence was identified as \texttt{65A8F92620B870000000C8C4}, where \texttt{0x26} (hexadecimal) represents the ID of the transmitting satellite. Although an offical, up-to-date mapping between the ID and ORBCOMM FM serial number is not publicly available, referring to the existing actual observation records \footnote{https://github.com/fbieberly/ORBCOMM-receiver} ( where ID \texttt{0x2C}/decimal 44 corresponds to ORBCOMM FM 114) suggests a consistent linear mapping between these parameters. The demodulated ID value (\texttt{0x26}/decimal 38) matches perfectly with ORBCOMM FM 108. This alignment not only substantiates our previous identification based on maximum declination but also provides empirical validation for the accuracy of our satellite transit prediction methodology.

\section{Discussion and Conclusions}
\label{conclusion}
In this work,  we conduct a preliminary search for unintended electromagnetic radiation (UEMR) from Starlink satellites using the 21CMA. Leveraging Two Line Elements (TLEs) data and the \texttt{Skyfield} Python library, we established a robust strategy to predict satellite transits within the 21CMA field of view. The reliability of this observational framework was subsequently validated through the successful demodulation of ORBCOMM downlink signals using our custom-developed \texttt{orbdemod} Python package. Our analysis of the system equivalent flux density (SEFD) and the sensitivity limits for a single pod observation provides an explanation for the absence of detectable UEMR in this experiment, pinpointing the current sensitivity threshold as the primary constraint. Furthermore, by employing modulation power spectrum analysis, we effectively excluded transient broadband bursts (with $34~\mu\text{s}$ duration) as UEMR candidates, confirming them to be power line arcing RFI. Using these events, we also performed a preliminary test of the feasibility of SAM2 for automated RFI identification. This result provides important practical guidance for mitigating terrestrial interference in future searches for satellite emission signals. 

Looking ahead, we plan to implement multi-beam coherent beamforming techniques for the 21CMA, which will substantially improve observational sensitivity and enable more definitive detection of potential Starlink UEMR. With this enhanced capability, we will extend our investigations to characterize the low-frequency emission properties of other emerging satellite constellations, including the Chinese Qianfan (G60) system. We will also examine whether measurable differences exist in the radiation signatures of Starlink satellites as they transit regions without active commercial service coverage. Such comparative analyses will help clarify the physical origins of UEMR and provide deeper insight into how operational modes or service configurations influence emission behavior. Ultimately, these studies will contribute to assessing the long-term impact of large satellite constellations on the radio astronomical environment.

\normalem
\begin{acknowledgements}

This work was funded by the National SKA Program of China (Grant No. 2020SKA0110200). YY acknowledges the support of the Key Program of the National Natural Science Foundation of China (12433012).

\end{acknowledgements}

\bibliographystyle{raa}
\bibliography{bibtex}

\end{document}